\documentclass[12pt]{article}
\usepackage{graphicx}
\usepackage{subcaption}
\usepackage[justification=centering]{caption}
\usepackage{wrapfig}

\usepackage{mathtools}
\usepackage{amssymb}
\usepackage{amsfonts}
\usepackage{float}
\usepackage[font=scriptsize]{subcaption}
\usepackage{booktabs}
\usepackage[labelfont=bf,textfont=md]{caption}
\usepackage{scrextend}
\deffootnote[2em]{2em}{1em}{\textsuperscript{\thefootnotemark}\,}

\usepackage[T1]{fontenc}

\newcommand\pubnumber{\hfill NuPhys2018-MolinaSedgwick}
\newcommand\pubdate{\hfill April 2019}

\def\napoli{$^{\star}$ School of Physics and Astronomy\\
University of Southampton, UK\\
-\\
$^{\dagger}$ Particle Physics Research Centre\\
Queen Mary University of London, UK\\
-\\
$^{\ddagger}$ Fermi National Accelerator Laboratory\\
Batavia (IL), USA\\
-\\
$^{\S}$ TRIUMF, Vancouver, British Columbia, Canada}
\def\support{\footnote{\textbf{NuPhys Speaker}\\
s.molina-sedgwick@soton.ac.uk, s.molinasedgwick@qmul.ac.uk}}

\def\Title#1{\begin{center} {\Large #1 } \end{center}}
\def\Author#1{\begin{center}{ \large \sc #1} \end{center}}
\def\Collaborators#1{\begin{center}{ \small #1}\end{center}}
\def\Address#1{\begin{center}{ \it #1} \end{center}}

\newcommand\pubblock{\rightline{\begin{tabular}{l} \pubnumber\\
         \pubdate  \end{tabular}}}
\newenvironment{Abstract}{\begin{quotation}  }{\end{quotation}}
\newenvironment{Presented}{\begin{quotation} \begin{center} 
             PRESENTED AT\end{center}\bigskip 
      \begin{center}\begin{large}}{\end{large}\end{center} \end{quotation}}
\def\Acknowledgements{\bigskip  \bigskip \begin{center} \begin{large}
             \bf ACKNOWLEDGEMENTS \end{large}\end{center}}

\def\beq{\begin{equation}}
\def\eeq#1{\label{#1}\end{equation}}
\def\eeqn{\end{equation}}

\def\beqa{\begin{eqnarray}}
\def\eeqa#1{\label{#1}\end{eqnarray}}
\def\eeqan{\end{eqnarray}}

\let\bar=\overbar

\def\Dslash{\not{\hbox{\kern-4pt $D$}}}
\def\dslash{\not{\hbox{\kern-2pt $\del$}}}

\def\msb{{\bar{\ssstyle M \kern -1pt S}}}

\begin{document}
\begin{titlepage}
\pubblock

\vfill
\Title{Matter density profile effects on neutrino\\
\vspace{0.2cm}
oscillations at T2HK and T2HKK}
\vfill
\Author{\textbf{Susana Molina Sedgwick}$^{\star,\dagger}$\support,\\
S.\,F.\,King$^{\star}$, S.\,J.\,Parke$^{\ddagger}$, N.\,W.\,Prouse$^{\S}$}
\vspace{-.75cm}
\Collaborators{}
\Address{\napoli}
\vfill
\begin{Abstract}
This project aims to explore the effects that changes in a matter density profile could have on neutrino oscillations, and whether these could potentially be seen by the future Hyper-Kamiokande experiment (T2HK). The analysis is extended to include the possibility of having a second detector in Korea (T2HKK).
\end{Abstract}
\vfill
\begin{Presented}
NuPhys2018, Prospects in Neutrino Physics\\
Cavendish Conference Centre, London, UK\\
\vspace{0.125cm}
December 19--21, 2018
\end{Presented}
\vfill
\end{titlepage}
\def\thefootnote{\fnsymbol{footnote}}
\setcounter{footnote}{0}

\section{Introduction}\

Following the success of the Kamiokande and Super-Kamiokande experiments, Hyper-Kamiokande will be the next-generation water Cherenkov detector in Japan. With a fiducial volume 10x greater than that of Super-K (or 20x greater if two tanks are built), Hyper-K will have a huge multipurpose research potential. It will be capable of studying everything from solar and atmospheric neutrinos to supernovae, as well as having applications to DM searches, neutrino tomography and proton decay. \cite{HK} \\

The long-baseline aspect of the experiment will be known as T2HK, and will involve a neutrino beam originating at J-PARC, Tokai, as well as an intermediate water Cherenkov detector and upgraded near detector. While this would have a baseline of approximately 300 km, it is a natural extension to consider placing any second tank further down the beamline, in South Korea, in order to be comparable to other current and future long-baseline experiments, such as NOvA and DUNE. \cite{HKK} \\

It is at these baselines that matter effects can become significant. As neutrinos propagate through the Earth, their oscillation probability is affected by the interactions between them and the protons, neutrons and electrons in matter. While neutrinos of all flavours experience neutral-current interactions with protons and neutrons, only electron neutrinos can undergo charged-current interactions with electrons, thus producing an asymmetry which is dependent on the number of electrons in matter - this in turn can be directly translated into matter density.\\

\begin{figure}[htb]
\centering
	\includegraphics[width=0.7\textwidth]{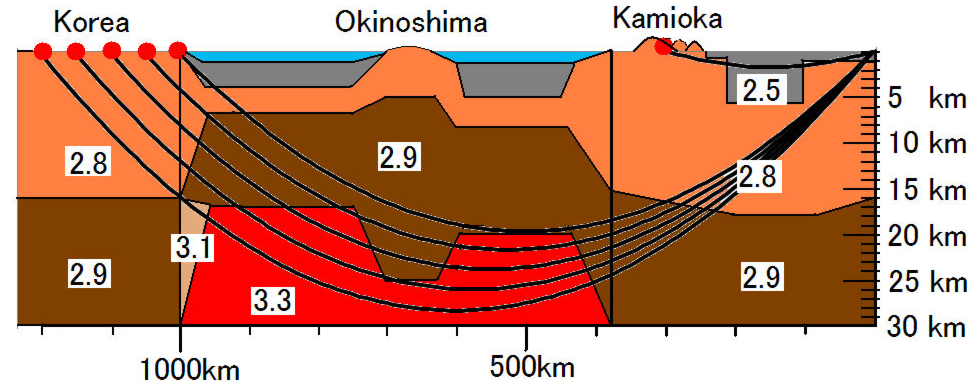}
\caption{Matter density profile of the Earth for T2HK and T2HKK.}\
\label{fig:Baselines}
\end{figure}

\textbf{Figure \ref{fig:Baselines}} shows a cross-section view of the matter density profile of the Earth between Tokai and Korea, with an approximate neutrino beamline for both Hyper-K and the 5 possible sites for a Korean detector (with baselines ranging from 1000 km to 1200 km). \cite{Hagiwara} \\

\vspace{-0.1cm}

\

While in the case of DUNE, it was concluded that varying the density profile would have no measurable effect on oscillation probabilities \cite{DUNE}, in this project we aim to determine whether the geographical features present in the beamlines for T2HK and T2HKK could produce a different result. To do this, we simulate these long-baseline experiments using the GLoBES program \cite{Globes1,Globes2}. It is important to note that for each result, only the appearance channel ($\nu_\mu\rightarrow\nu_e$) is shown here.


\section{Sensitivity Estimates}\
\label{Sensitivity}

Using GLoBES, a $1\sigma$ region of allowed oscillation parameters (based on a $\chi^2$ calculation) is converted into an uncertainty on the oscillation probability itself. These results are obtained from just 100 energy steps and 100 steps in each of the two parameters being scanned over, in order to keep computational time low. Future calculations will improve on this, and an additional method of estimating the uncertainty will be explored so as to compare the two alternatives.\\

\begin{figure}[htb]
\centering
	\includegraphics[width=0.495\textwidth]{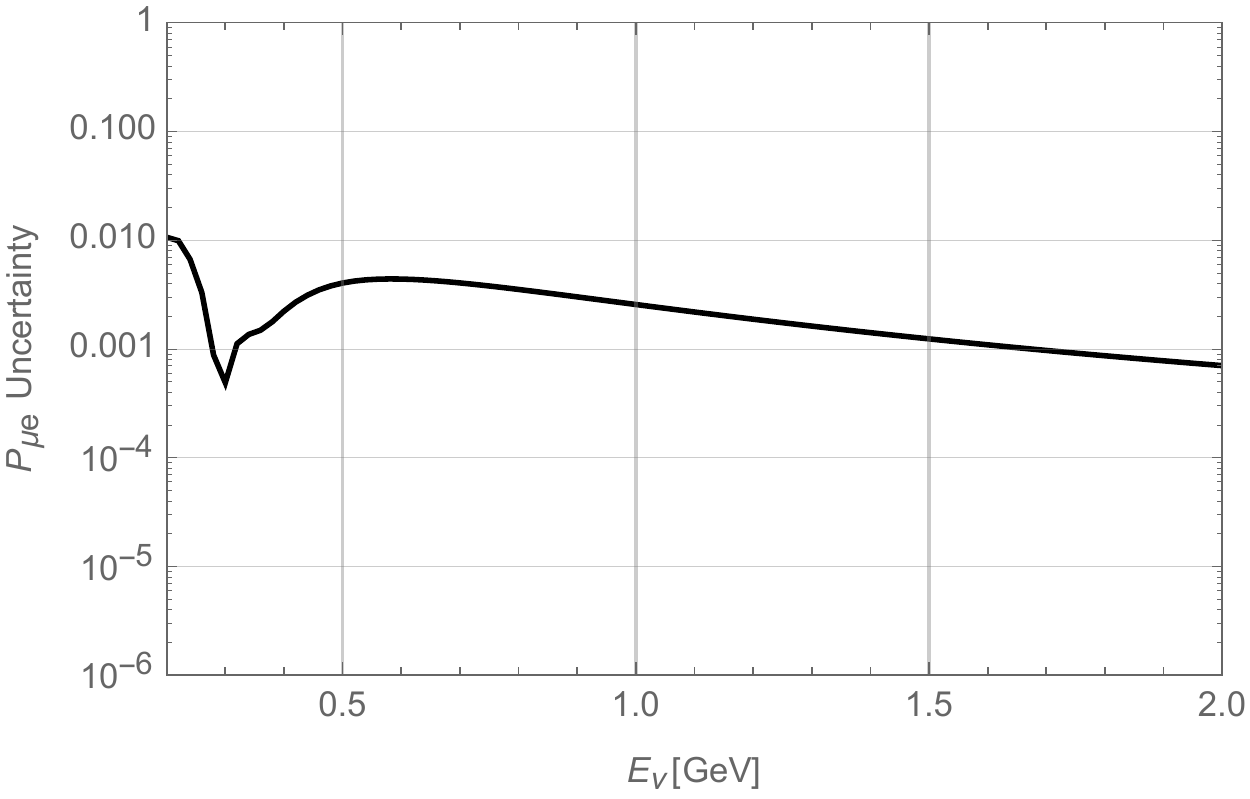}
	\includegraphics[width=0.495\textwidth]{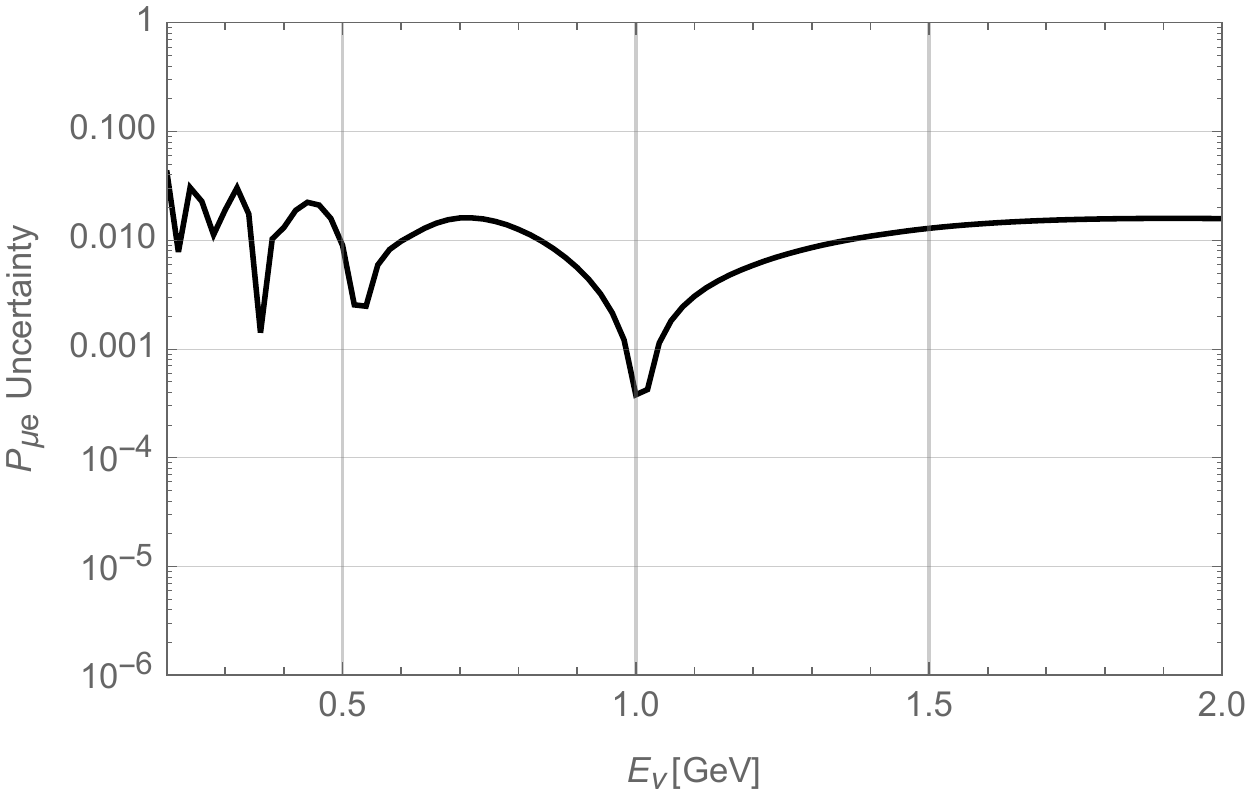}
\caption{Estimate of uncertainty on oscillation probability\\
for Hyper-K (left) and Hyper-K Korea (right).}\
\label{fig:Sensitivities}
\end{figure}

\vspace{-0.75cm}

\section{Parameter Variation}\

\begin{figure}[htb]
\centering
	\includegraphics[width=0.495\textwidth]{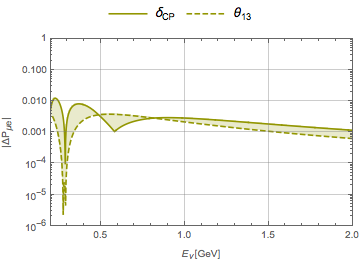}
	\includegraphics[width=0.495\textwidth]{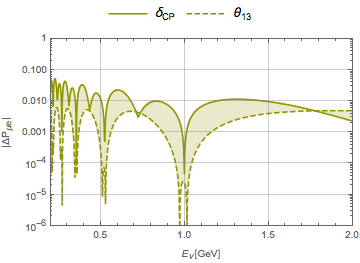}
\caption{Changes in oscillation probability due to a variation of $\delta_{CP}$ (solid)\\
and $\theta_{13}$ (dashed) for Hyper-K (left) and Hyper-K Korea (right).}\
\label{fig:ParamScan_Var}
\end{figure}

In \textbf{Figure \ref{fig:ParamScan_Var}}, we show the changes in neutrino oscillation probability arising from a variation of the oscillation parameters between their $+1\sigma$ and $-1\sigma$ values. These values have been taken from the Hyper-K Design Report \cite{HK}, with the exception of $\theta_{13}$, which is given by the current NuFit range. \cite{NuFit} \\

At each energy point, we scan over the entire range between the two values (100 steps) and plot the difference between the maximum and minimum calculated probability, rather than simply assuming that e.g. the maximum probability will correspond to the maximum value of a specific parameter - this takes into account possible non-linear contributions. Again, only the parameters measured by the appearance channel are shown here ($\delta_{CP}$ and $\theta_{13}$).\\


\vspace{-0.75cm}

\section{Matter Density Profile Variation}
\label{Density}

\begin{figure}[htb]
\centering
	\includegraphics[width=0.495\textwidth]{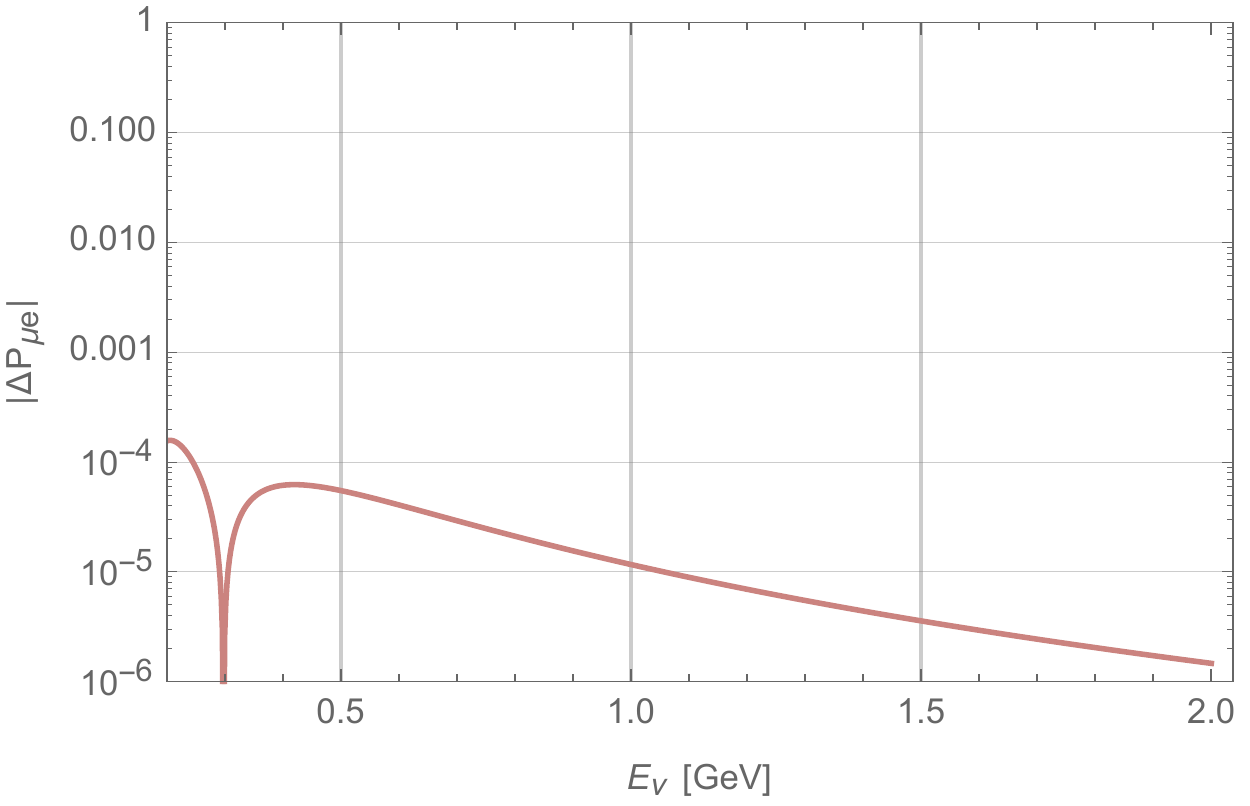}
	\includegraphics[width=0.495\textwidth]{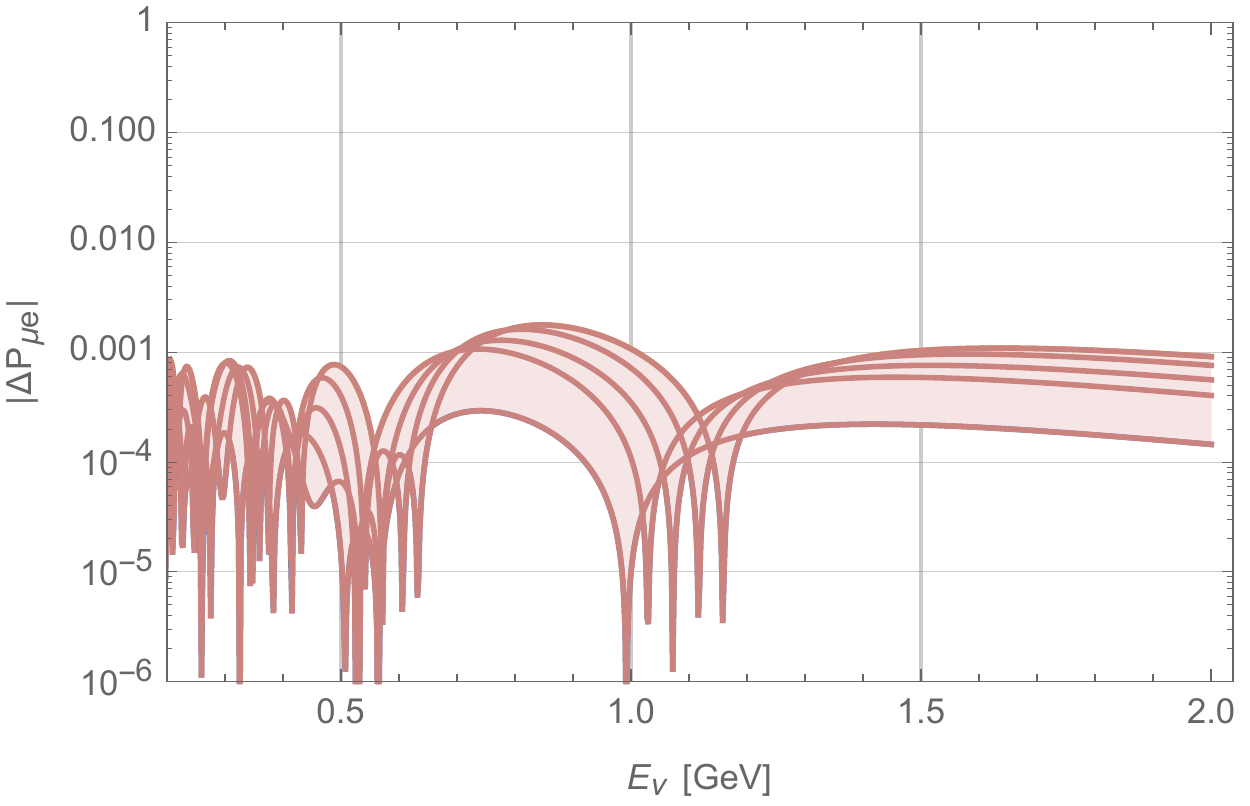}
	\caption{Changes in oscillation probability with respect to average density at Hyper-K (left) and the different baselines of Hyper-K Korea (right).}\
\label{fig:Densities}
\end{figure}

\textbf{Figure \ref{fig:Densities}} shows the changes in oscillation probability as a result of changes in the matter density profile. The curves are produced by taking the difference between the probability at each point calculated using the varying (realistic) matter density profile and using a constant density. For the constant profile, the average value of the varying density was used - in the case of Hyper-K Korea, the changes in oscillation probability of each individual baseline profile
 are shown on the right.\\

\section{Additional and Future Studies}\

We performed two additional studies. First, we looked at the changes in oscillation probability given by varying the average density of each baseline profile by $\pm 1\%$, in order to compare them to the results from \textbf{Section \ref{Density}}.\\

\begin{wrapfigure}{r}{0.495\textwidth}
\centering
	\includegraphics[width=0.495\textwidth]{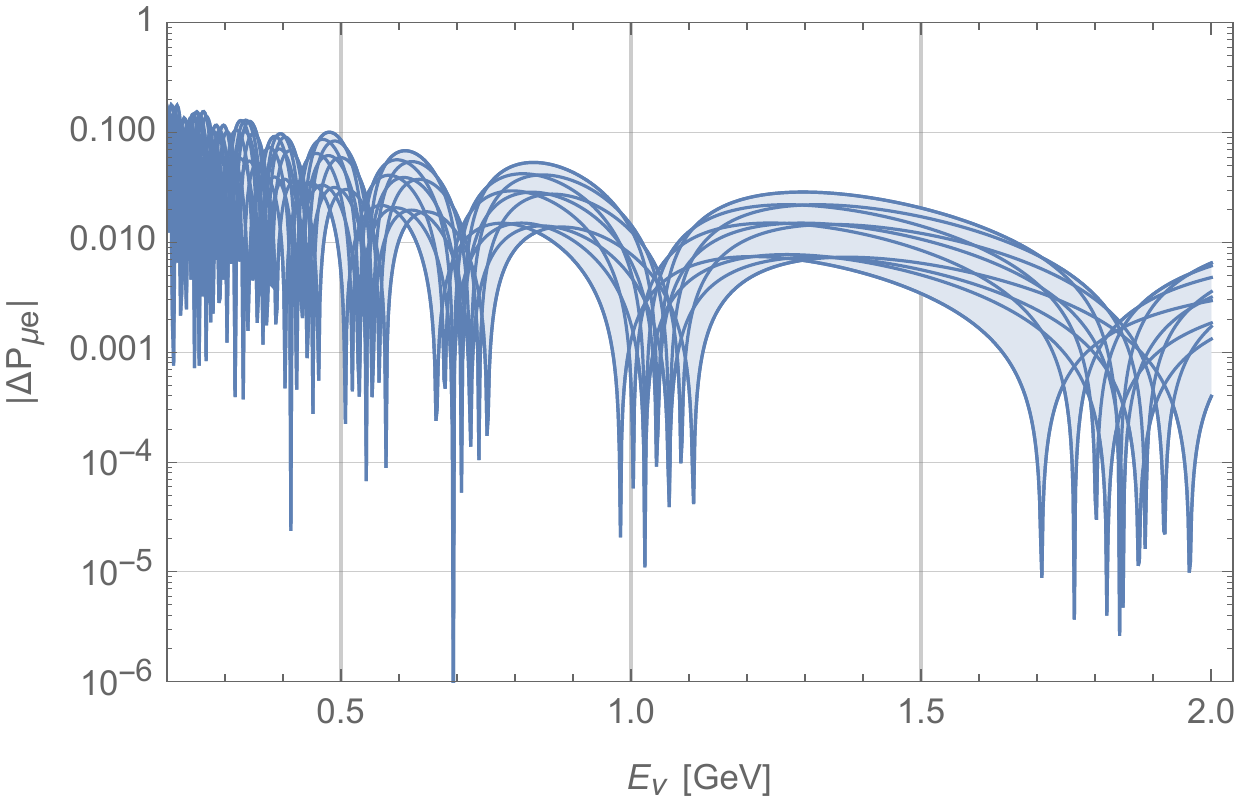}
	\caption{Changes in oscillation probability from the differences between baselines at Hyper-K Korea.}\
\label{fig:Differences}
\end{wrapfigure}\

\vspace{-0.1cm}

In addition to that, we explored the changes in probability arising from the differences between each of the 5 baseline profiles from Hyper-K Korea. These results are shown in \textbf{Figure \ref{fig:Differences}} - there are ten curves, each corresponding to the difference of probabilities between two baseline profiles. It can be seen that this is the only test that produces a result above the estimated sensitivity line from \textbf{Section \ref{Sensitivity}}. This hints at the possibility of the effects of changing from one profile to another being potentially detectable at T2HKK.\\

Although these are purely preliminary results, it can be clearly seen that there are significant differences between matter effects in T2HK and T2HKK. Moreover, we can conclude that a varying matter density profile at T2HKK could potentially give rise to detectable effects on neutrino oscillation probabilities.\\

In order to probe this effect further, we aim to use an alternative method of approximating the effect of different density profiles - by focusing on the most likely candidate site for the Korean detector, we can redo the calculations after removing each of the density ``chunks" in turn. This can give us a better idea for how each of the different density areas under the Earth's surface will affect the final result.

\pagebreak

\Acknowledgements\

I would like to thank my collaborators S.\,F.\,King, S.\,J.\,Parke and N.\,W.\,Prouse for their guidance and support during this project.\\

This work was made possible in part due to an InvisiblesPlus Network secondment to Fermilab in May 2018. I wish to acknowledge the UK Science and Technology Facilities Council (STFC) Doctoral Training Grant held by Queen Mary University of London, with additional support from the University of Southampton and the Valerie Myerscough Science and Mathematics Trust Fund at the University of London.

\end{document}